\documentclass[a4paper,11pt]{article}
\usepackage{amssymb,amsmath,bm}  % for math
\usepackage{graphics,graphicx}   % for figures
\usepackage{array,booktabs}      % for tables
\usepackage{authblk}  % for footnote style author/affiliation
\usepackage{slashed}

\usepackage{cite}
\usepackage{xcolor}
\usepackage[margin=2.5cm]{geometry}

\usepackage[linktocpage,
colorlinks = true,
linkcolor = blue,
urlcolor  = blue,
citecolor = red,
anchorcolor = blue]{hyperref}

\usepackage{epsfig}

\numberwithin{equation}{section}

\numberwithin{equation}{section}

\title{\textbf{Gravitational wave in a filtered vector dark matter model}}
\author[a]{Mojtaba Hosseini \thanks{mojtaba\textunderscore hosseini@semnan.ac.ir }}
\author[a]{Seyed Yaser Ayazi\thanks{syaser.ayazi@semnan.ac.ir}}
\author[b]{Ahmad Mohamadnejad\thanks{mohamadnejad.a@lu.ac.ir}}
\affil[a]{Physics Department, Semnan University, P.O. Box. 35131-19111, Semnan, Iran}
\affil[b]{Department of Physics, Lorestan University, Khorramabad, Iran}

\date{\today}
\begin{document}

\baselineskip 0.6 cm
\maketitle

\begin{abstract}
We consider a first order phase transition (FOPT) for a Vector Dark Matter (VDM) in the early universe in which its mass may partially arise from such mechanism in the hidden sector. We calculate the ratio of VDM that may
enter the bubble for various bubble wall velocities as well as various nucleation temperatures that produce the measured dark matter relic abundance via bubble filtering. In the following, we focus on gravitational wave (GW) signals which produced by FOPT and show that it can be detectable at the  DECIGO,  TianQin,  LISA and Ultimate-DECIGO (UDECIGO) experiments.
\end{abstract}

%\newpage

%\tableofcontents

\section{Introduction} \label{sec1}
There is a lot of evidence that indicates that about 27\% of the content of our universe is made of an unknown substance called dark matter (DM)\cite{Bertone:2016nfn}. The absence of a suitable candidate for DM in the Standard Model (SM) leads us to build models beyond the SM. There are different methods to detect DM such as direct detection and indirect detection. In this regard, many experiments have been designed to find a clue of DM through direct or indirect search, but so far no evidence of DM has been obtained.

In addition, the identity of DM and its production mechanism are open questions in particle physics. Among the various mechanism, the weakly interacting massive particles (WIMPs) with freeze-out production are the most popular models\cite{Feng:2022rxt}. WIMPs have been the main target of many DM direct detection experiments. None of the results of direct detection experiments so far have found any sign of DM. For this reason, the existence of DM with very large masses can be important. There is an upper bound on the WIMP mass at about 100 TeV\cite{Griest:1989wd,Baldes:2017gzw,Smirnov:2019ngs},
above which the WIMP interaction required by the observed relic density violates unitarity. This upper bound known as the Griest-Kamionkowski (GK) bound. Different methods beyond the traditional freeze-out mechanism have been provided to escape this constraint(GK bound)\cite{Kolb:2017jvz,Kim:2019udq}. Our approach in this article is the filtering-out effect during the FOPT\cite{Baker:2019ndr,Chway:2019kft,Chao:2020adk}.

It is possible that the DM mass is not a constant, but is dynamically generated from spontaneous breaking of symmetry such as Higgs mechanism
or chiral symmetry breaking by strong dynamics. Then,
when the Universe is hot enough, thermal effects restore
the symmetry prohibiting the DM mass. As the temperature drops below the critical temperature, phase transition begins and DM gains nonzero mass.
During the bubble expansion, the potential energy stored in the symmetric vacuum is converted mostly into the bulk kinetic energy of the plasma fluid surrounding the bubble wall. It is because the fluid
pressure is equilibrated with the potential difference.
When bubbles collide, most of the bulk kinetic energy is
converted into the thermal energy. If the corresponding phase transition is first order, bubbles of the broken phase nucleate and expand during the phase transition. Since the symmetry is unbroken outside bubbles, the DM is still massless there, while inside the
bubbles the symmetry is broken, resulting in a nonzero
mass of DM.
The massless DM particles that exist outside the bubbles undergo a transformation into massive particles once they enter the bubbles. However, it is important to note that only the massless DM particles carrying a kinetic energy greater than $ m_{DM} $ are able to penetrate the walls of the bubbles and acquire mass. On the other hand, the DM particles residing within the bubbles experience a sudden decoupling from the thermal bath if $ T_n < T_{dec} $, where $ T_n $ ($ T_{dec} \approx \frac{m_{DM}}{24} $) is the nucleation temperature (the decoupling temperature). As a consequence, the bubbles act as filters, causing a certain amount of DM to be filtered out and ultimately determining the relic abundance of DM. It is worth mentioning that the massless DM particles that are located outside the bubbles continue to be thermally connected to the SM radiation. Furthermore, DM particles lacking sufficient kinetic energy to enter the bubbles will be reflected back into the symmetric phase, thereby slowing down the expansion of the bubbles as they exert pressure on the bubble walls.
In this paper, we follow the idea of Refs. \cite{Baker:2019ndr,Chway:2019kft} and
investigate the filtering-out effect of a vector DM. We study GW signal induced by the FOPT at the space-based interferometer such as LISA\cite{Caprini:2015zlo}, DECIGO\cite{Seto:2001qf}, Ultimate-DECIGO(UDECIGO)\cite{Yagi:2011wg} and TianQin\cite{TianQin:2015yph}. Gravitational waves can be a specific probe for the filtering-out mechanism. The interesting aspect of the paper is that the findings concerning DM density primarily rely on hydrodynamic rather than being heavily influenced by the specific DM model. However, the spectrum of gravitational waves is contingent upon the potential and differs from the results obtained through other models.

The remaining of the paper is organized as follows: In section~\ref{sec2}, we introduce the model. In section~\ref{sec3}, Bubble filtering and its effect on VDM are presented.  In section~\ref{sec4}, we study direct detection of VDM in XENONnT.
Section~\ref{sec5} is focused on the investigation of GW signals at the space-based
interferometer and we calculate the GW signals for the
two benchmark of the model. We summarize in section~\ref{sec6}.

\section{The Model} \label{sec2}
In the model, beyond the SM, we employ two new fields \cite{YaserAyazi:2019caf}: a  scalar field $S$ which has unit charge under a dark $ U'(1) $ gauge symmetry with a dark photon vector field $ V_{\mu} $. The model has an additional $Z_2$ discrete symmetry, under which the vector field $ V_{\mu} $ and the scalar field transform as follows: $V_{\mu} \rightarrow - V_{\mu}$, $S\rightarrow S^*$ and all the other fields are even.  $Z_2$ symmetry forbids the kinetic mixing between the  the vector field $ V_{\mu} $ and SM $ U_{Y}(1) $ gauge boson $ B_{\mu} $, i.e., $ V_{\mu \nu} B_{\mu \nu} $. Therefore,  the vector field $ V_{\mu} $ is stable and can be considered as a DM candidate. The Lagrangian can be written as:
\begin{equation}
 {\cal L} ={\cal L}_{SM} + (D'_{\mu} S)^{*} (D'^{\mu} S) - V(H,S) - \frac{1}{4} V_{\mu \nu} V^{\mu \nu} , \label{2-2}
\end{equation}
where $ {\cal L} _{SM} $ is the SM Lagrangian without the Higgs potential term and
\begin{align}
& D'_{\mu} S= (\partial_{\mu} + i g_vV_{\mu}) S,\nonumber \\
& V_{\mu \nu}= \partial_{\mu} V_{\nu} - \partial_{\nu} V_{\mu},\nonumber \end{align}
and the most general scale-invariant potential which is renormalizable and invariant
under gauge and $ Z_{2} $ symmetry is:
\begin{equation}
V(H,S) = \frac{\lambda_{H}}{6} (H^{\dagger}H)^{2} +\frac{\lambda_{S}}{6}  (S^*S)^{2} + 2 \lambda_{S H} (S^*S) (H^{\dagger}H). \label{2-3}
\end{equation}

Note that the quartic portal interaction, $ \lambda_{SH} (S^*S) (H^{\dagger}H) $, is the only connection between the dark sector and the SM.

SM Higgs field $ H $ as well as dark scalar $S$ can receive VEVs breaking respectively the electroweak and $ U'_{D}(1) $ symmetries.
In unitary gauge, the imaginary component of $S$ can be absorbed as the longitudinal component of $ V_{\mu} $.
In this gauge, we can write:
\begin{equation}
H = \frac{1}{\sqrt{2}} \begin{pmatrix}
0 \\ h_{1} \end{pmatrix} \, \, \, and \, \, \, S = \frac{1}{\sqrt{2}} h_{2} , \label{2-4}
\end{equation}
where $ h_{1} $ and $ h_{2} $ are real scalar fields which can get VEVs.
The tree level potential in unitary gauge is given by:
\begin{equation}
V_{\text{tree}}(h_{1},h_{2})=\frac{1}{4!} \lambda _H h_1^4 +\frac{1}{4!} \lambda _S h_2^4+\frac{1}{2} \lambda _{SH} h_1^2 h_2^2.\label{2-4}
\end{equation}
Vacuum stability requires $\lambda _{H,S}>0$ and $ \lambda _{SH}<0$. Furthermore, non-zero
VEV of $h_{1,2}$ scalar fields demands $\lambda _H \lambda _S=(3!\lambda _{SH})^2$.

The Local minimum of the two-variable potential (\ref{2-4})
defines a direction in field-space known as flat direction\cite{Gildener:1976ih}. Along this direction $V_{tree} = 0$, while in other directions $V_{tree} > 0$. The full potential of the theory will be dominated by higher-loop
contributions along flat direction and specifically by the one-loop effective potential. Indeed, for some mass spectrum of the
model, 1-loop effective potential, $ V_{eff}^{1-loop} $, gives a small curvature in the flat direction with $V_{eff}^{1-loop} < 0$. Since at the minimum of the one-loop
effective potential $V_{tree}\geq 0$ and $V_{eff}^{1-loop} < 0$, the minimum of $V_{eff}^{1-loop}$ along the flat direction
(where $V_{tree} = 0$) is a global minimum of the full potential, therefore spontaneous symmetry
breaking occurs and we should substitute $h_1 \longrightarrow \nu_1 + h_1$ and $h_2 \longrightarrow \nu_2 + h_2$. We suppose $\nu_1$ and $\nu_2$ are
VEVs of $h_1$ and $h_2$ where $\nu_1 = 246$ GeV. The mass eigenstates $H_{1}$ and $H_{2}$ can be rewritten in term of  $h_{1}$ and $h_{2}$:
\begin{equation}
\begin{pmatrix}
H_{1}\\H_{2}\end{pmatrix}
=\begin{pmatrix} cos \alpha~~~  -sin \alpha \\sin \alpha  ~~~~~cos \alpha
\end{pmatrix}\begin{pmatrix}
h_1 \\  h_{2}
\end{pmatrix}, \label{2-11}
\end{equation}
where $H_{2}$ is along the flat direction, thus $ m_{H_{2}} = 0 $, and $ H_{1} $ is perpendicular to the flat direction which we identify it as the SM-like Higgs observed at the LHC with $ m_{H_{1}} = 125 $ GeV. After the symmetry breaking, we have the following constraints:
\begin{align}
& \nu_{2} =  \frac{m_{V}}{g_v} , &\nonumber
& sin \alpha =  \frac{\nu_{1}}{\sqrt{\nu_{1}^{2}+\nu_{2}^{2}}} \nonumber \\
& \lambda_{H} =  \frac{3 m_{H_{1}}^{2}}{ \nu_{1}^{2}} cos^{2} \alpha  \nonumber&
& \lambda_{S} =  \frac{3 m_{H_{1}}^{2}}{ \nu_{2}^{2}} sin^{2} \alpha  &\nonumber\\
& \lambda_{S H} =  - \frac{ m_{H_{1}}^{2}}{2 \nu_{1} \nu_{2} } sin \alpha \, cos \alpha , \label{2-6}
\end{align}
where $m_V$ is the mass of VDM after symmetry breaking.

In tree level, the scalon field $H_2$ is massless, however the radiative corrections including the one-loop
corrections to the potential give a mass to the massless eigenstate $H_2$ (Gildener-Weinberg mechanism\cite{Gildener:1976ih}). Regarding 1-loop effect, the scalon mass is given by\cite{YaserAyazi:2019caf}
\begin{equation}
m_{H_{2}}^{2} = \frac{1}{8 \pi^{2} \nu^{2}} \left( m_{H_{1}}^{4} + 6  m_{W}^{4} + 3  m_{Z}^{4} + 3  m_{V}^{4} - 12 m_{t}^{4} \right)\label{2-7},
\end{equation}
where $ m_{W,Z,t}$ are the masses of W, Z gauge bosons, and top quark, respectively. From (\ref{2-7}), we have a constraint on the parameter space of the model where  $ m_V>240$ GeV. Constraints (\ref{2-6}) severely restrict free parameters of the model up to two independent parameters. We choose $m_V$ and $g_v$ as the independent parameters of the model.

\section{Bubble Filtering-out effect} \label{sec3}
The mass gap between outside and inside of the bubble plays the key role in the filtering-out mechanism. If the energy
of a small mass DM particle outside the bubble is smaller
than the energy gap, it cannot enter the bubble because of
the energy conservation. DM particles that do not have
enough energy are filtered out.

In this section, we study the consequence of FOPT followed by bubble dynamics and find a new application of filtering-out effect for VDM. In the model, $V_{\mu}$ partially acquires a mass from the phase transition in the dark sector.

A FOPT is processed by the bubble nucleation of the true vacuum. DM particles $V_{\mu}$ initially are massless and in thermal equilibrium with SM particles and scalar particle $S$. Only DM particles with enough momentum to overcome their mass inside the bubbles can pass through the walls. Otherwise, the DM particles are reflected and stays outside the bubble. So, we assume the bubble wall moves in the negative z direction with velocity $v_{\omega}$ and during the FOPT, VDM with momentum $p = (p_x , p_y , p_ z )$ in wall frame  can penetrate into the wall only when its z-direction momentum in wall frame satisfy this condition:
\begin{equation}
\gamma_{\omega}(p_z+v_{\omega}E_{\omega})>\sqrt{\Delta m^2}
\end{equation}
where $v_{\omega}$, $\gamma_{\omega}=1/{\sqrt{1-v^2_\omega}}$ and $E_{\omega}=\sqrt{|p|^2+m_0^2}$ are wall velocity, Lorentz boost factor  and energy in wall frame, respectively. $\Delta m^2$ is equal to the mass difference of $V_{\mu}$ between the outside of the bubble and after the phase transition.

To clearly see that the hydrodynamic effects play important roles in the VDM relic density, we consider analytic estimation of the VDM density. The detailed numerical calculation of the general DM density by using the Boltzmann equations to include the hydrodynamic effects have been studied in \cite{Baker:2019ndr}. It was also shown that, those numerical results  are in agreement with the analytical estimates.  Therefore, in order to calculate the relic abundance, we follow  analytical estimation which have been presented in\cite{Marfatia:2020bcs}.

We start with the Maxwell-Boltzmann approximation for the equilibrium distribution function, and consider the Bose-Einstein distribution for VDM:
\begin{equation}
		f^{eq}_V=\frac{1}{e^{\gamma_{\omega}(E_P/T_n)}-1}
\end{equation}
	where $T_n$ and $E_p$ are the bubble nucleation temperature and the energy in plasma frame, respectively. The energy in the plasma is given by:
\begin{equation}
		E_P=\gamma_{\omega}(E_{\omega}-v_{\omega}p_z)
\end{equation}
	where $p$ is momentum in the wall's rest frame. We expect
	interactions with the wall to cause deviations from equilibrium. The particle flux arising from the false vacuum per unit area and unit time can be written as\cite{Chway:2019kft}
\begin{align}
		\tilde{J_V}&=G_V\int \!\! \frac{d^3\tilde{\textbf{p}}}{(2\pi)^3}\frac{\tilde p_z}{\tilde E} f_V\Theta \left(\tilde p_z-\sqrt{\Delta m_{}^2 }\right) \; \label{xxxx}
\end{align}	where $G_V$ =3 is the number of spin states of the VDM. The VDM number density inside the bubble $n_V^{\rm in}$ in plasma rest frame can be given by~\cite{Chway:2019kft}:
\begin{align}
		n_V^{\rm in}&=\frac{\tilde{J_V}}{\gamma_{w}v_w} \; .
		\label{number density}
\end{align}
We can integrate \ref{xxxx} and get the number density of VDM\cite{Chway:2019kft,Marfatia:2020bcs}:
\begin{eqnarray}
	\label{eq:filter_analytic}
	n^{\rm in}_V\simeq\frac{G_VT^3_n}{\gamma_w v_w}\left(\frac
	{\gamma_w (1-v_w)m_V /T_n+1}{4 \pi^2 \gamma^3_\omega(1-v_w)^2}
	\right)e^{-\frac{\gamma_w (1-v_w)m_\chi}{T_n}}\,.
\end{eqnarray}
where we have used Maxwell-Boltzmann approximation of DM distribution.
In the non-relativistic limit, $v_w \to 0$, filtering strongly suppresses
the DM number density inside the bubble as $e^{-m_V/T_n}$.
In the relativistic limit, $m_V/(\gamma_w T_n)\to 0$, the number density
$\sim e^{-m_V/(2 \gamma_w T_n)}$, so there is very little filtering and
$n^{\rm in}_V$
approaches the equilibrium number density outside the bubble, $n^{\rm eq}_{\chi}|_{T=T_n} = G_VT^3_n/\pi^2$.

The VDM abundance today can be calculated by scaling inside number of VDM density $n_V^{\rm in}$ at $T_n$ with the entropy density $s=(2\pi^2/45)g_{\star S}T^3$ and the critical density, $\rho_c = 3H^2_0 m^2_{\rm pl}$ ($m_{\rm pl}$ is the reduced Planck mass,  $H_0 = 100 \, h \ {\rm km} / \text{sec} / \text{Mpc}$ is the Hubble constant, $g_{\star S} \sim 106.75$ at $T_n$ and $g_{\star S}\equiv g_{\star S0}\sim 3.9$  today). We can get the current DM relic abundance in relativistic bubble wall velocities limit ($v_w\rightarrow 1$) with this relation\cite{Marfatia:2020bcs,Baker:2019ndr,Chway:2019kft}:
\begin{align}
\label{omega}
\Omega_{\rm DM}h^2
&\simeq 1.27\times 10^8
\left( \frac{m_V}{\rm GeV} \right)
\left( \frac{G_V}{g_{\star S}} \right)
\left(1+\frac{m_V}{2\gamma_{\omega}T_n} \right)e^{-\frac{m_V}{2\gamma_{\omega}T_n}}.
\end{align}
 In non-relativistic bubble wall velocities limit ($v_w\rightarrow 0$), we obtain\cite{Marfatia:2020bcs,Baker:2019ndr,Chway:2019kft}:
\begin{align}
\label{omeganon}
\Omega_{\rm DM}h^2
&\simeq 3.19\times 10^7
\left( \frac{m_V}{\rm GeV} \right)
\left( \frac{G_V}{g_{\star S}} \right)
\left( \frac{1}{v_w} \right)
\left(1+\frac{m_V}{T_n} \right)e^{-\frac{m_V}{T_n}}.
\end{align}
Figure \ref{Omega1} shows relic density as a function of $m_V/T_n$ for non-relativistic and relativistic bubble wall velocities.

Also, Figure \ref{Omega2} shows the allowed points in the $m_V -T_n$ plane in agreement with the relic density of VDM\cite{Planck:2018vyg}. As shown in Figure \ref{Omega2}, in order to meet the relic abundance bound, smaller values of $m_V/T_n$ are required in the non-relativistic limit compared to the relativistic limit. This means that the filtering effect on relativistic bubble wall velocities is minimal, whereas the filtering effect on non-relativistic bubble wall velocities can be significant. This also shows that the allowed range in the $m_V -T_n$ plane are very narrow due to the exponential dependence of relic density to $m_V/T_n$.
{The results of our calculations about the VDM show a similar pattern about the toy model presented in \cite{Baker:2019ndr} and other models, because the density of DM depends only on the mass of DM, number of degrees of freedom  and $T_n$. But depending on the type of DM particle and the number of its spin states, the allowed parameter space of the model will be different. However, what distinguishes the remarkable predictions of different models in terms of phenomenology is the production of GWs.  Because different potentials are used to calculate the spectrum of GWs and this issue shows itself in the final results.

\begin{figure}%[!htb]
\begin{center}
\centerline{\hspace{0cm}\epsfig{figure=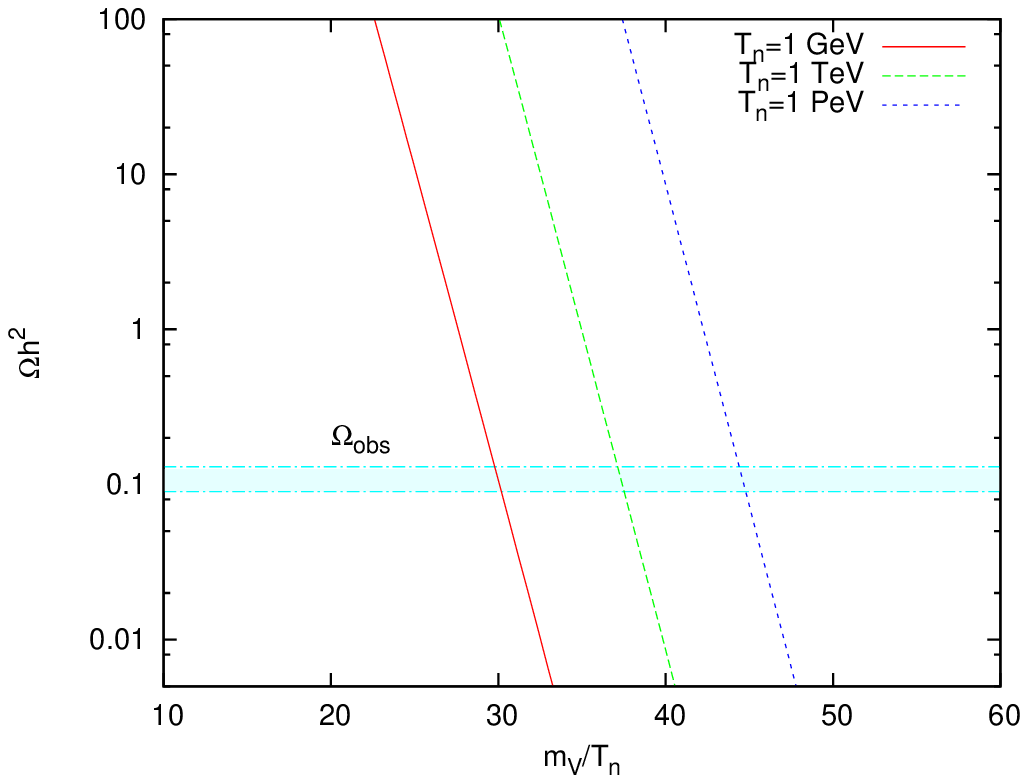,width=7.5cm}\hspace{0cm}\epsfig{figure=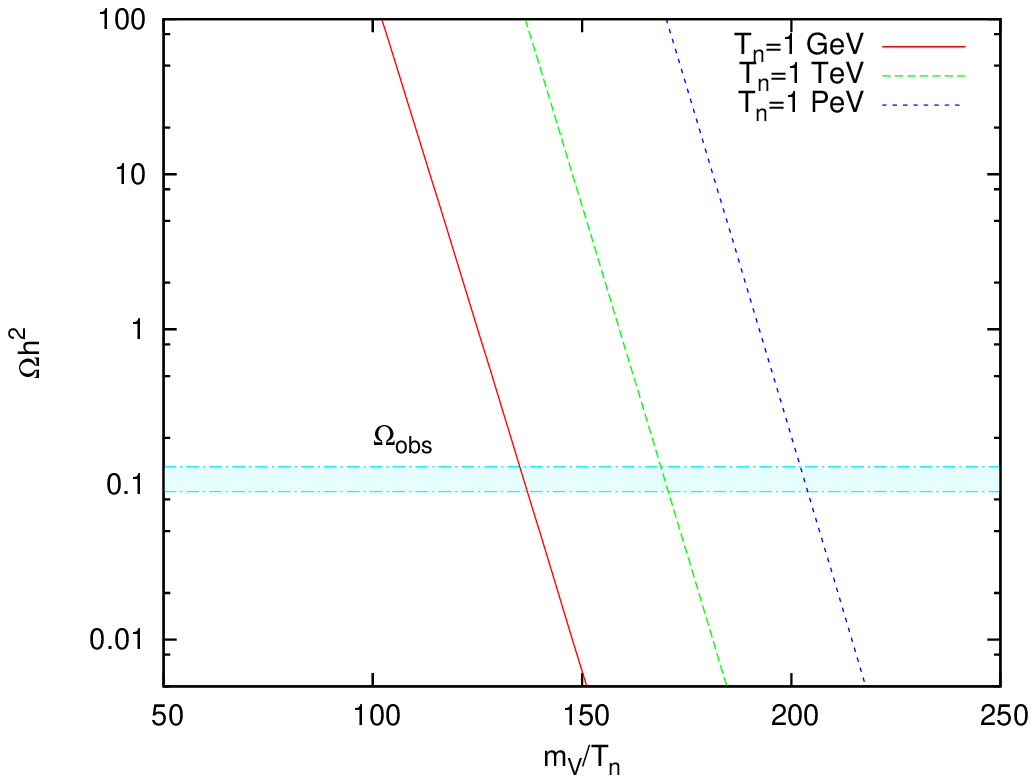,width=7.5cm}}
\centerline{\vspace{0.5cm}\hspace{0.5cm}(a)\hspace{6cm}(b)}
\centerline{\vspace{-0.7cm}}
\caption{The relic density as a function of $m_V/T_n$ for a) non-relativistic $v_w=0.01$ and b) relativistic $v_w=0.9$	bubble wall velocities.}\label{Omega1}
\end{center}
\end{figure}

\begin{figure}%[!htb]
	\begin{center}
		\centerline{\hspace{0cm}\epsfig{figure=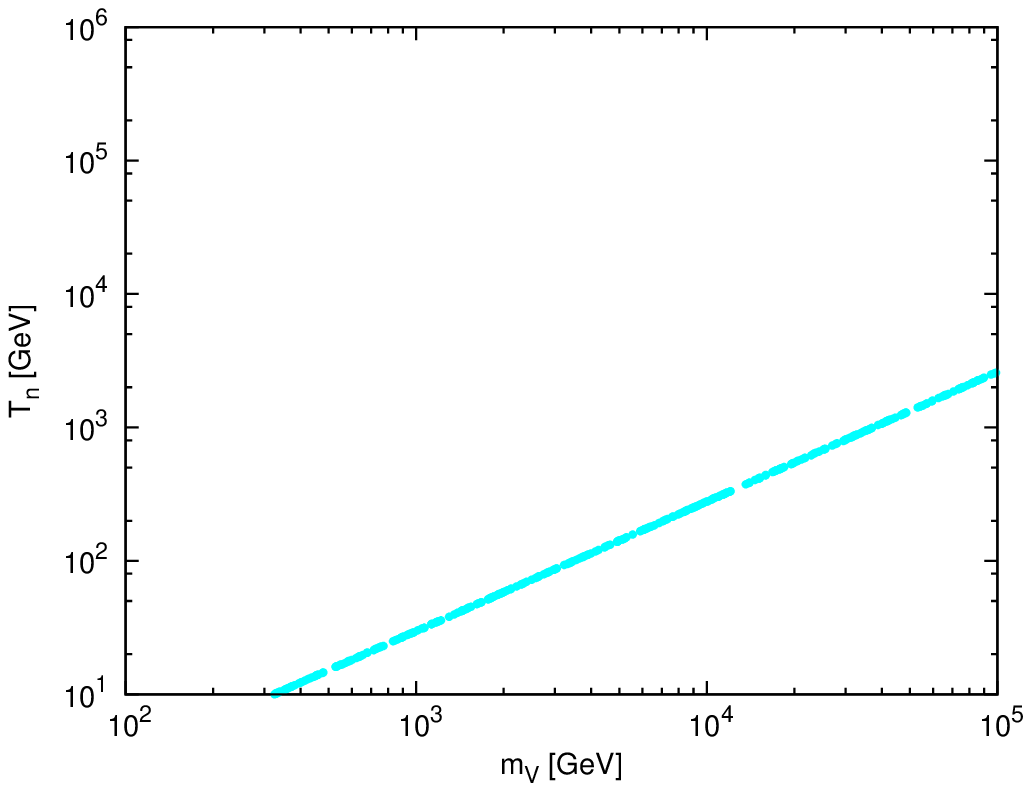,width=7.5cm}\hspace{0cm}\epsfig{figure=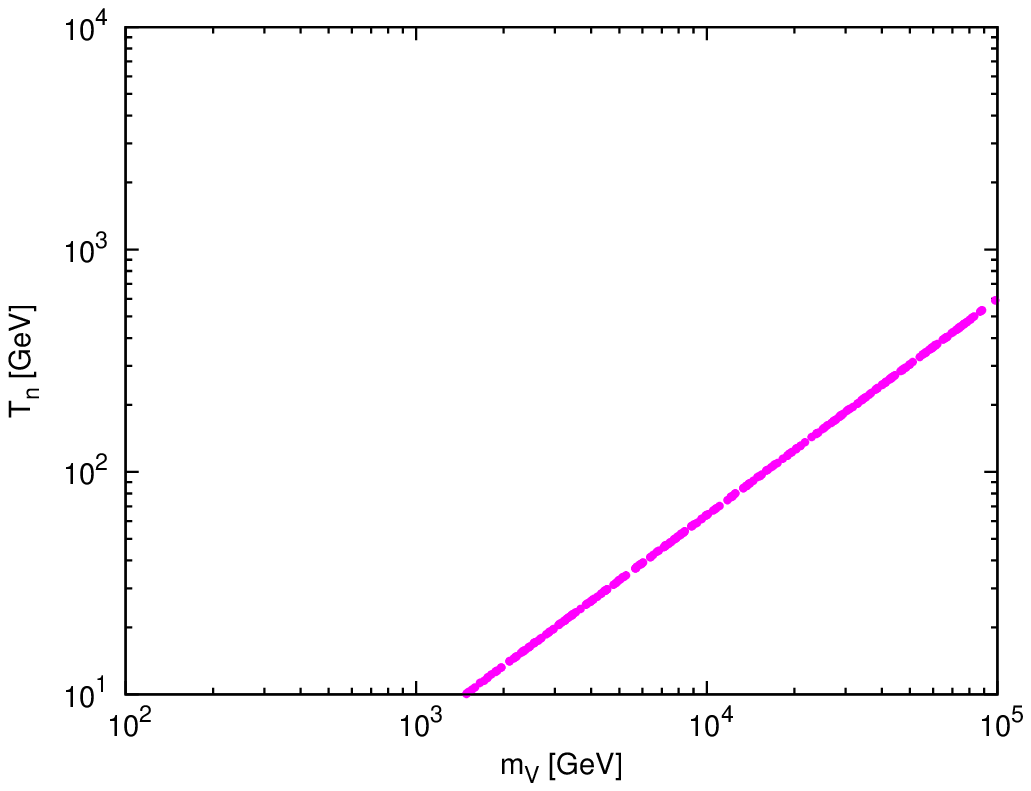,width=7.5cm}}
		\centerline{\vspace{0.5cm}\hspace{0.5cm}(a)\hspace{6cm}(b)}
		\centerline{\vspace{-0.7cm}}
		\caption{The scatter points depict allowed ranges in $m_V$ and $T_n$ plane for a) non-relativistic $v_w=0.01$ and b) relativistic $v_w=0.9$	bubble wall velocities. }\label{Omega2}
	\end{center}
\end{figure}

\section{Direct Detection}\label{sec4}
Direct detection of DM particles is mediated, in this model, via exchange of $H_2$ particles and
Higgs bosons ($H_1$). The spin-independent direct detection (DD) cross section of $V$ is determined by $H_1$ and $H_2$ exchanged
diagrams\cite{YaserAyazi:2019caf}:
\begin{equation}\label{DD}
	\sigma ^{DM-N} = \frac{4 \lambda_{SH} ^2 m_V ^2 m_N ^2 \mu_{VN} ^2 (m_{H1} ^2 -m_{H2} ^2 )^2}{\pi m_{H1} ^8 m_{H2} ^4}f_N ^2 ,
\end{equation}
where
\begin{equation}
	\mu_{VN} = m_N m_V / (m_N + m_V).
\end{equation}
$m_N$ is the nucleon mass and $f_N\simeq 0.3$ parametrizes the Higgs-nucleon coupling.

We probe the proposed model and mechanism (Filtered DM) with experimental results. We have introduced two benchmark points in Table \ref{table2}. These points satisfy the observed relic density. Figure \ref{cross section} shows the $\sigma ^{DM-N}$ for these points. We use the XENONnT\cite{XENON:2023cxc} experiment results. The neutrino floor is also shown which is a the irreducible background coming from scattering of SM neutrinos on nucleons\cite{Billard:2013qya}. As it is known, for masses of smaller than 800~$\rm GeV$, DD is not possible, but for masses of larger than that, DD is still possible. Even for very large masses, where the figure is below the neutrino floor, gravitational waves can be a specific probe for the model\cite{Mohamadnejad:2019vzg,Hosseini:2023qwu}, and this condition can be important for the Filtered DM mechanism. Both of the benchmarks are consistent with experimental results.

We also used constraints related to mixing between scalars ($sin \alpha$) in checking our results. Non-observation of DM at the LHC corresponds to an upper bound on the value of $sin \alpha$. The Higgs boson data from LHC at 7 and 8 TeV sets a constraint of $|sin \alpha| < 0.36$ at 95\% C.L., independently of the
$H_2$ mass\cite{Carena:2018vpt}. There is also a mass-dependent constraint,
which requires $sin \alpha < 0.32$ for $m_{H2}> 200 GeV$ and $sin \alpha < 0.2$ for $m_{H2}> 400 GeV$, mostly
coming from restrictions on the NLO corrections to the mass of the W boson\cite{Robens:2015gla}.

\begin{figure}%[!htb]
	\begin{center}
		\centerline{\hspace{0cm}\epsfig{figure=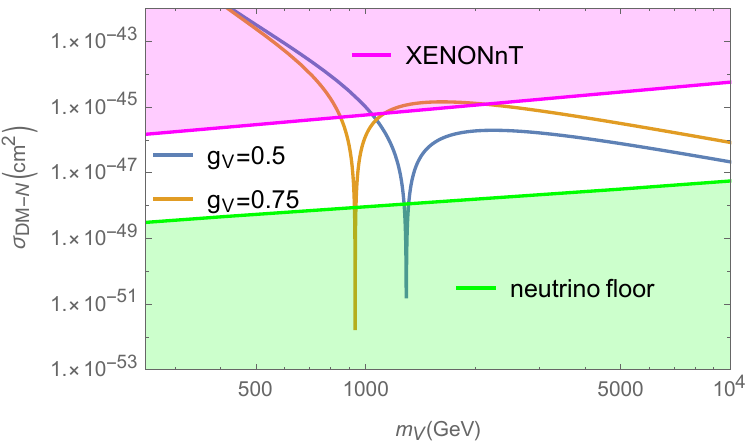,width=12cm}}
		\centerline{\vspace{-0.2cm}}
		\caption{ DM-nucleon cross section as a function of VDM mass for differnet values of
			coupling $g_v$.} \label{cross section}
	\end{center}
\end{figure}

\section{Gravitational Wave Signals} \label{sec5}
The transition from the false to the true vacuum proceeds via thermal tunneling
at finite temperature. Once this has
happened, the bubble spreads throughout the Universe converting false vacuum
into true one. The bubbles formation starts at the nucleation temperature $T_n$, where one can
 estimates $T_n$ by the condition $S_3 (T_n) / T_n \sim 140$\cite{Apreda:2001us}. The function $S_3(T)$ is the three-dimensional Euclidean
action for a spherical symmetric bubble given by
\begin{equation}\label{Euclidean action}
S_3(T)= 4\pi \int_{0}^{\infty} dr  r^2 \Biggl(\frac{1}{2} \Bigl(\frac{dH_2}{dr}\Bigl)^2 +V_{eff}(H_2 ,T)\Biggl),
\end{equation}
where $H_2$ satisfies the differential equation which minimizes $S_3$:
\begin{equation}\label{S3}
\frac{d^2 H_2}{dr^2} + \frac{2}{r} \frac{dH_2}{dr}=\frac{dV_{eff}(H_2 ,T)}{dH_2},
\end{equation}
with the boundary conditions:
\begin{equation}
\left. \frac{dH_2}{dr} \right|_{r=0}=0, ~~~~ and ~~~~ H_2 (r\longrightarrow \infty )=0.
\end{equation}
In order to solve Eq. \ref{S3} and find the Euclidean action \ref{Euclidean action}, we used the AnyBubble
package\cite{Masoumi:2017trx}.

GWs resulting from the strong first-order electroweak phase transitions are caused by three contributions, which are as follows:

$\bullet$ collisions of bubble walls and shocks in the plasma,

$\bullet$ sound waves to the stochastic background after collision of bubbles but before expansion

 has dissipated the kinetic energy in the plasma

 $\bullet$ turbulence forming after bubble collisions.

 These three processes may coexist, and each one contributes to the stochastic GW background:
\begin{equation}
\Omega_{GW} h^{2} \simeq \Omega_{coll} h^{2}+\Omega_{sw} h^{2}+\Omega_{turb} h^{2}.
\end{equation}
There are four thermal parameters that control the above contributions:

 $\bullet$ $T_n$:  the nucleation temperature,

  $\bullet$ $\alpha$: the ratio of the free energy density difference
between the true and false vacuum and

 the total energy density,
\begin{equation}
\alpha= \frac{\Delta \Bigl(V_{eff} -T\frac{\partial V_{eff}}{\partial T}\Bigl)\bigg\vert_{T_n}}{\rho_\ast},
\end{equation}

where $\rho_\ast$ is
\begin{equation}
\rho_\ast= \frac{\pi^2 g_\ast}{30}T_n^4,
\end{equation}
 $\bullet$ $\beta$:  the inverse time duration of the phase transition,
 \begin{equation}
\frac{\beta}{H_\ast}= T_n \frac{d}{dT}\Bigl(\frac{S_3 (T)}{T}\Bigl)\bigg\vert_{T_n},
\end{equation}

$\bullet$ $\upsilon_\omega$: the velocity of the bubble wall.

The collision contribution to the spectrum is
given by\cite{Huber:2008hg}
\begin{equation}
\Omega_{coll}(f) h^{2}=1.67\times 10^{-5} \Bigl(\frac{\beta}{H_\ast}\Bigl)^{-2} \Bigl(\frac{\kappa \alpha}{1+\alpha} \Bigl)^2 \Bigl(\frac{g_\ast}{100} \Bigl)^{-\frac{1}{3}} \Bigl(\frac{0.11 \upsilon_\omega^3}{0.42+\upsilon_\omega^2}\Bigl) S_{coll},
\end{equation}
where $S_{coll}$ parametrises the spectral shape and is given by
\begin{equation}
S_{coll}=\frac{3.8 (f/f_{coll})^{2.8}}{2.8 (f/f_{coll})^{3.8} +1 },
\end{equation}
where
\begin{equation}
f_{coll}= 1.65\times 10^{-5} \Bigl(\frac{0.62}{\upsilon_\omega^2 -0.1\upsilon_\omega +1.8}\Bigl)  \Bigl(\frac{\beta}{H_\ast}\Bigl) \Bigl(\frac{T_n}{100}\Bigl) \Bigl(\frac{g_\ast}{100} \Bigl)^{1/6} Hz.
\end{equation}

The collision of bubbles produces a massive movement in the fluid in the form of sound waves that generate GWs. This
is the dominant contribution to the GW signal which is given by\cite{Hindmarsh:2015qta}
\begin{equation}
\Omega_{sw}(f) h^{2}=2.65\times 10^{-6} \Bigl(\frac{\beta}{H_\ast}\Bigl)^{-1} \Bigl(\frac{\kappa_\upsilon \alpha}{1+\alpha} \Bigl)^2 \Bigl(\frac{g_\ast}{100} \Bigl)^{-\frac{1}{3}} \upsilon_\omega S_{sw}.
\end{equation}
The spectral shape of $S_{sw}$ is
\begin{equation}
S_{sw}= (f/f_{sw})^3 \Bigl(\frac{7}{3 (f/f_{sw})^2 +4} \Bigl)^{3.5},
\end{equation}
where
\begin{equation}
f_{sw}= 1.9\times 10^{-5} \frac{1}{\upsilon_\omega} \Bigl(\frac{\beta}{H_\ast}\Bigl) \Bigl(\frac{T_n}{100}\Bigl) \Bigl(\frac{g_\ast}{100} \Bigl)^{1/6} Hz.
\end{equation}

Plasma turbulence can also be caused by bubble collisions, which is a contributing factor to the GW spectrum and is given by\cite{Caprini:2009yp}
\begin{equation}
\Omega_{turb}(f) h^{2}=3.35\times 10^{-4} \Bigl(\frac{\beta}{H_\ast}\Bigl)^{-1} \Bigl(\frac{\kappa_{turb} \alpha}{1+\alpha} \Bigl)^{3/2} \Bigl(\frac{g_\ast}{100} \Bigl)^{-\frac{1}{3}}  \upsilon_\omega S_{turb},
\end{equation}
where
\begin{equation}\label{Sturb}
S_{turb}= \frac{(f/f_{turb})^3}{(1+8\pi f/h_\ast)(1+f/f_{turb})^{11/3}},
\end{equation}
and
\begin{equation}
f_{turb}= 2.27\times 10^{-5} \frac{1}{\upsilon_\omega} \Bigl(\frac{\beta}{H_\ast}\Bigl) \Bigl(\frac{T_n}{100}\Bigl) \Bigl(\frac{g_\ast}{100} \Bigl)^{1/6} Hz.
\end{equation}

In Eq.~\ref{Sturb}, $h_\ast$ is the value of the inverse Hubble time at
GW production, redshifted to today,
\begin{equation}
h_\ast= 1.65\times 10^{-5} \Bigl(\frac{T_n}{100}\Bigl) \Bigl(\frac{g_\ast}{100} \Bigl)^{1/6}.
\end{equation}

For computing GW spectrum, we have used\cite{Caprini:2015zlo,Kamionkowski:1993fg}
\begin{align}
& \kappa= \frac{1}{1+0.715\alpha}(0.715\alpha + \frac{4}{27}\sqrt{\frac{3\alpha}{2}}) , \nonumber \\
& \kappa_\upsilon= \frac{\alpha}{0.73 + 0.083\sqrt{\alpha}+\alpha},~~~~ \kappa_{turb}=0.05\kappa_\upsilon ,
\end{align}
where the parameters $\kappa$, $\kappa_\upsilon$, and $\kappa_{turb}$ denote the fraction of latent heat that is transformed into gradient energy of the Higgs-like field, bulk motion of the fluid, and MHD turbulence, respectively.

\begin{table}[h]
\centering % centering table
\begin{tabular}{l c c rrrrrrr} % creating 10 columns
\hline\hline
 $\#$ &$m_V (GeV)$ &$g_v$&$\sigma_{DM-N} (cm^2)$&$T_n (GeV)$&$\alpha$&$\beta/H_\ast$&$v_w$\\
\hline
1&1184&0.75&$8.38\times10^{-46}$&34.5&100.07&51.11&0.01 \\
2&4000&0.5&$1.09\times10^{-46}$&112.5&115.39&549.08&0.01\\

\end{tabular}
\caption{\label{table2}Two benchmark points with DM and phase transition parameters.} % title name of the table
\end{table}

To investigate the GWs resulting from the first-order  electroweak phase transition, we need to find points whose nucleation temperature ($T_n$) can be obtained from the presented model potential and which is within the observed relic density range. Due to the limited parameter space in figure \ref{Omega2}, it is very difficult to find these points. we choose two benchmark points.  These points are shown in Table \ref{table2} and are in agreement with figure \ref{Omega2}. In both of our benchmarks, a strong filtering effect ($v_w\rightarrow 0$) has been chosen to investigate gravitational waves. To clarify the phase transition parameters, we have presented an example of our analysis process in figure \ref{critical} for Benchmark 1. In figure \ref{critical} potential behavior is given for critical temperature ($T_C$) and nucleation temperature ($T_n$). Also $S_3/T$ changes in terms of temperature is given for  benchmark 1. The GW spectrum for these benchmark points is depicted in figure \ref{GW2}. The GW spectrum for these benchmarks 1,(2) falls within the observational window of LISA, DECIGO, UDECIGO and TianQin(DECIGO and UDECIGO).

\begin{figure}%[!htb]
\begin{center}
\centerline{\hspace{0cm}\epsfig{figure=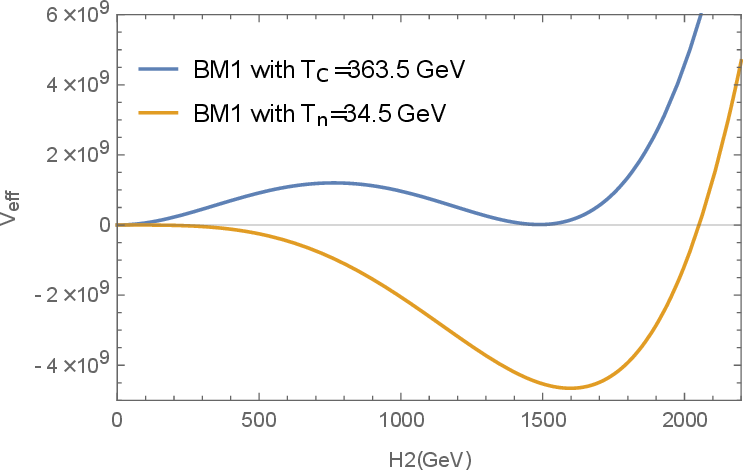,width=7.5cm}\hspace{0cm}\epsfig{figure=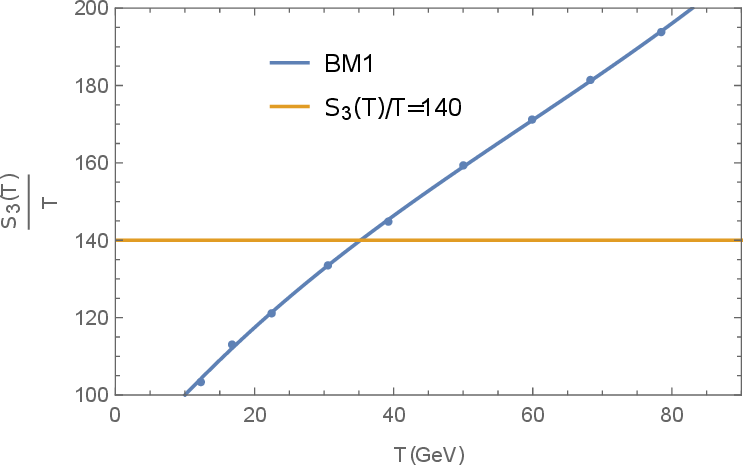,width=7.5cm}}
\centerline{\vspace{0.5cm}\hspace{0.5cm}(a)\hspace{6cm}(b)}
\centerline{\vspace{-0.7cm}}
\caption{In (a) Potential behavior is given for critical temperature and nucleation temperature. In (b), $S_3/T$ changes in terms of temperature is also given for  benchmark 1.}\label{critical}
\end{center}
\end{figure}

\begin{figure}%[!htb]
	\begin{center}
		\centerline{\hspace{0cm}\epsfig{figure=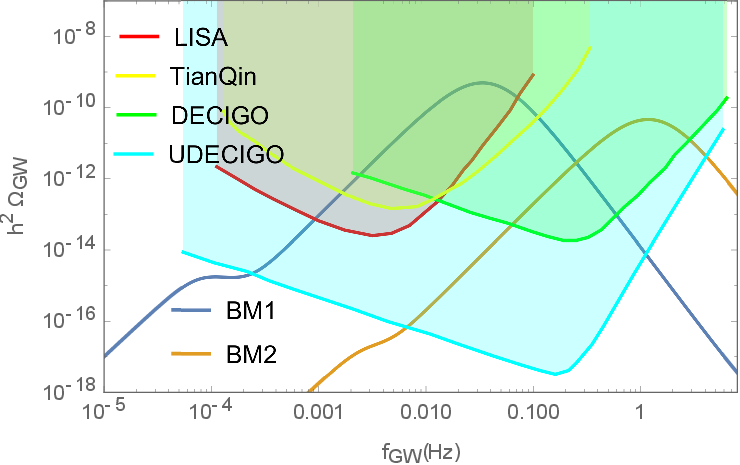,width=12cm}}
		\centerline{\vspace{-0.2cm}}
		\caption{GW spectrum for benchmark points of the table \ref{table2}.} \label{GW2}
	\end{center}
\end{figure}

\section{Conclusion} \label{sec6}
We have studied VDM and its dynamics during a FOPT in the early universe. The study explored how VDM interacted during this critical period, particularly emphasizing the filtering-out effect where VDM particles were selectively removed from regions inside the bubbles formed during the phase transition. This effect had implications for the distribution of DM and its detectability in experiments like XENONnT.

Furthermore, we have studied the GW signals generated by the phase transition, highlighting the potential for detecting primordial GWs using space-based interferometers such as LISA, DECIGO, TianQin, and UDECIGO. By analyzing these signals, we aimed to refine the parameters of the VDM model and test its predictions against observational data.

The study presented two benchmark points that aligned with observed DM relic density and expected GW signals, serving as examples of how the VDM model could successfully account for these phenomena in the early universe. The discussion also mentioned limitations related to scalar mixing and other experimental results.

In summary, this study highlighted the close connection between DM, phase transitions, and GWs, emphasizing the potential of VDM as a reliable DM candidate.
 By merging theoretical calculations with observational constraints, the study aimed to offer profound insights into physics beyond the SM and illuminate the enigmatic nature of DM.

\section*{Appendix: effective potential}
 Along the flat direction, the one-loop effective potential, has the general form \cite{Gildener:1976ih}
\begin{equation}
V_{eff}^{1-loop(T=0)} = a H_{2}^{4} + b H_{2}^{4} \, \ln \frac{H_{2}^{2}}{\Lambda^{2}}  , \label{2-13}
\end{equation}
where  $ a $ and $ b $ are the dimensionless constants that given by
\begin{align}
& a =  \frac{1}{64 \pi^{2} \nu^{4}}  \sum_{k=1}^{n} g_{k}  m_{k}^{4} \ln \frac{m_{k}^{2}}{\nu^{2}}  , \nonumber \\
& b = \frac{1}{64 \pi^{2} \nu^{4}} \sum_{k=1}^{n} g_{k}  m_{k}^{4} . \label{2-14}
\end{align}
In (\ref{2-14}), $ m_{k} $ and $ g_{k} $ are, the  tree-level mass and the internal degrees of freedom of the particle $ k $, respectively (In our convention $ g_{k} $
takes positive values for bosons and negative ones for
fermions).
By minimizing the relation \ref{2-13} and rewriting in terms of the one-loop VEV $\nu$, we have
\begin{equation}
V_{eff}^{1-loop(T=0)} = b H_{2}^{4} \, \left( \ln \frac{H_{2}^{2}}{\nu^{2}} - \frac{1}{2} \right) .\label{2-16}
\end{equation}
Regarding $ V_{eff}^{1-loop(T=0)} $, $ m_{H_{2}} $  will be
\begin{equation}
m_{H_{2}}^{2} = \frac{d^2 V_{eff}^{1-loop(T=0)}}{d H_{2}^{2}} \bigg\rvert_{\nu} = 8 b \nu^{2} , \label{2-17}
\end{equation}
Which leads to relation \ref{2-7}.

In addition to the 1-loop zero-temperature potential (\ref{2-16}), we can also consider the 1-loop corrections at finite
temperature in the effective potential, which is as follows\cite{Dolan:1973qd}
\begin{equation}\label{finite}
V_{eff}^{1-loop(T\neq0)}(H_{2},T) = \frac{T^4}{2\pi^2}\sum_{k=1}^{n} g_{k} J_{B,F} \Bigl(\frac{m_k}{\nu}\frac{H_2}{T}\Bigl) ,
\end{equation}
with thermal functions
\begin{equation}
J_{B,F}(x)= \int_{0}^{\infty} dy  y^2  ln \Bigl(1\mp e^{-\sqrt{y^2+x^2}}\Bigl).
\end{equation}
The above functions can be expanded in terms of modified Bessel functions
of the second kind, $K_2 (x)$\cite{Mohamadnejad:2019vzg},
\begin{align}
& J_B (x)\simeq -\sum_{k=1}^{3} \frac{1}{k^2}x^2 K_2 (kx) , \nonumber \\
& J_F (x)\simeq -\sum_{k=1}^{2} \frac{(-1)^k}{k^2}x^2 K_2 (kx). \nonumber \\
\end{align}
The contribution of resummed daisy graphs is also as follows\cite{Carrington:1991hz}
\begin{equation}\label{daisy}
V_{daisy}(H_2 ,T)=  \sum_{k=1}^{n} \frac{g_k T^4}{12\pi}  \Biggl(\Bigl(\frac{m_k}{\nu}\frac{H_2}{T}\Bigl)^3 - \biggl(\Bigl(\frac{m_k}{\nu}\frac{H_2}{T}\Bigl)^2 + \frac{\Pi_k (T)}{T^2}\biggl)^{\frac{3}{2}} \Biggl),
\end{equation}
where the sum runs only over scalar bosons and longitudinal degrees of freedom of the
gauge bosons. Thermal masses, $\Pi_k (T)$, are given by
\begin{align}
& \Pi_W= \frac{11}{6}g_{SM}^2 T^2 \nonumber, ~~~~~~~~~~ \Pi_V=\frac{2}{3}g_{v}^2 T^2\nonumber,~~~~~~~~ \Pi_{Z/\gamma}=\frac{11}{6}\begin{pmatrix}g_{SM}^2~~~~~~0\\0~~~~~~g_{SM}^{\prime2}\end{pmatrix}T^2,\\
&  \Pi_{H_1 /H_2}=\begin{pmatrix}\frac{\lambda_H}{24}+\frac{\lambda_{SH}}{12}+\frac{3g_{SM}^2}{16}+\frac{g_{SM}^{\prime2}}{16}+\frac{\lambda_{t}^2}{4}~~~~~~~~~0~~~~~\\~~~~~~~~~~~~~~~~~0~~~~~~~~~~~~~~~~~~~\frac{\lambda_S}{24}+\frac{\lambda_{SH}}{12}+\frac{g_{v}^2}{4}\end{pmatrix}T^2.
\end{align}

Finally, the one-loop effective potential  is given by
\begin{equation}\label{full potential}
V_{eff}(H_2 ,T)= V^{1-loop(T=0)}(H_2) +  V^{1-loop(T\neq0)}(H_2 ,T)+V_{daisy}(H_2 ,T)
\end{equation}
In order to get $V_{eff}(0 ,T)=0$ at all temperatures, We make the following substitution:
\begin{equation}
V_{eff}(H_2 ,T) \longrightarrow V_{eff}(H_2 ,T) - V_{eff}(0 ,T).
\end{equation}

\bibliography{References}
\bibliographystyle{JHEP}

\end{document}